\documentclass[prb,floatfix,aps,showpacs,superscriptaddress]{revtex4}
\usepackage[dvips]{graphics}
\usepackage[english]{babel}
\usepackage{graphicx}
\usepackage{amssymb}
\usepackage{epstopdf}
\usepackage{latexsym}
\usepackage{amsmath}
\usepackage{color}
\newcommand{\dn}{\downarrow}
\newcommand{\up}{\uparrow}
\newcommand{\Tr}{{\rm Tr}}

\begin{document}

\title{Tensor product variational formulation applied to pentagonal lattice}

\author{Michal Dani\v{s}ka and Andrej~Gendiar}
\affiliation{Institute of Physics, Slovak Academy of Sciences, SK-845~11, Bratislava, Slovakia}

\begin{abstract}
The uniform two-dimensional variational tensor product state is applied to the
transverse-field Ising, XY, and Heisenberg models on a regular hyperbolic lattice
surface. The lattice is constructed by tessellation of the congruent pentagons with
the fixed coordination number being four. As a benchmark, the three models are studied
on the flat square lattice simultaneously. The mean-field-like universality of the Ising
phase transition is observed in full agreement with its classical counterpart on the
hyperbolic lattice. The tensor product ground state in the thermodynamic limit has
an exceptional three-parameter solution. The variational ground-state energies of the
spin models are calculated.
\end{abstract}


\maketitle

\section{Introduction}

The tensor product state (TPS) has been proved to be an appropriate ansatz for
obtaining a ground-state of strongly correlated quantum systems. Numerous
computational approaches and methods have been developed and successfully applied to
one-dimensional (1D) and two-dimensional (2D) quantum systems~\cite{Orus,Uli,Scholl-RMP}.
The main purpose of this work is to address the tensor product state to two-dimensional
spin systems on a non-trivial lattice configuration, which is represented by an infinite
hyperbolic surface. The lattice is made from the pentagonal tessellation and forms
a negatively curved surface of the constant Gaussian curvature. Such a lattice geometry
has not yet been considered for 2D quantum systems in the thermodynamic limit. Lacking
information on constructing an appropriate numerical algorithm for the pentagonal lattice
is given.

The adherence to the mean-field-like universality class was concluded in the 2D classical
Ising systems on various types of hyperbolic lattices. The mean-field-like behavior of
the phase transition exponents is caused by the non-Euclidean geometry of the underlying
lattices~\cite{hctmrg-Ising-5-4,hctmrg-Ising-p-4,hctmrg-Ising-3-q}, and is not related
to the analysis by the mean-field approximation.
The classical spin systems have been analyzed by the Corner Transfer Matrix
Renormalization Group (CTMRG) method~\cite{ctmrg-tn}, which is a variant of
the DMRG for 1D quantum systems~\cite{dmrg-srw}. Now, we extend our earlier
studies, which dealt with the classical spin systems, to quantum spin systems
on the curved pentagonal lattice surface.

We, therefore, propose an approximative scheme of the {\it round-a-face} TPS, which
is related to the standard {\it vertex type} TPS with two-state auxiliary
variables~\cite{Uli,Orus}. Our main intention is to approximate the ground state
of spin-$\frac{1}{2}$ models by a minimal number of variational parameters. Improvements
of the numerical accuracy go beyond the scope of this work.

The paper is organized in the following way. Section II specifies the model Hamiltonians
on the Euclidean and the non-Euclidean lattices. The variational approach is discussed
with respect to the TPS approximation. A short description of the numerical algorithm and
discussion of tensor symmetries are included within two subsections. The numerical results
are analyzed in Sec.~III, and we comment the results in Sec.~IV.

\section{The Model}

\begin{figure}[tb]
\centerline{\includegraphics[width=0.23\textwidth,clip]{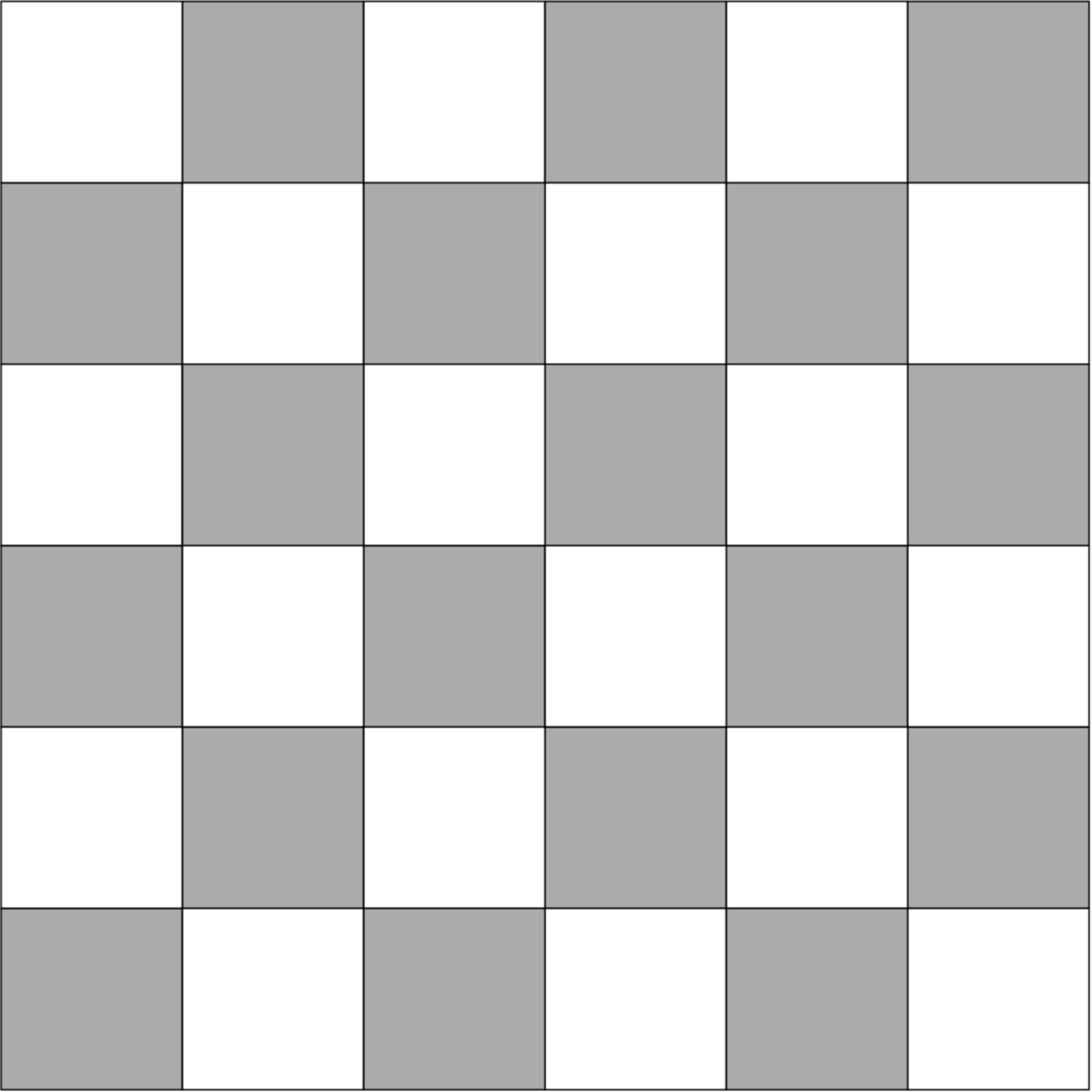}
            \includegraphics[width=0.25\textwidth,clip]{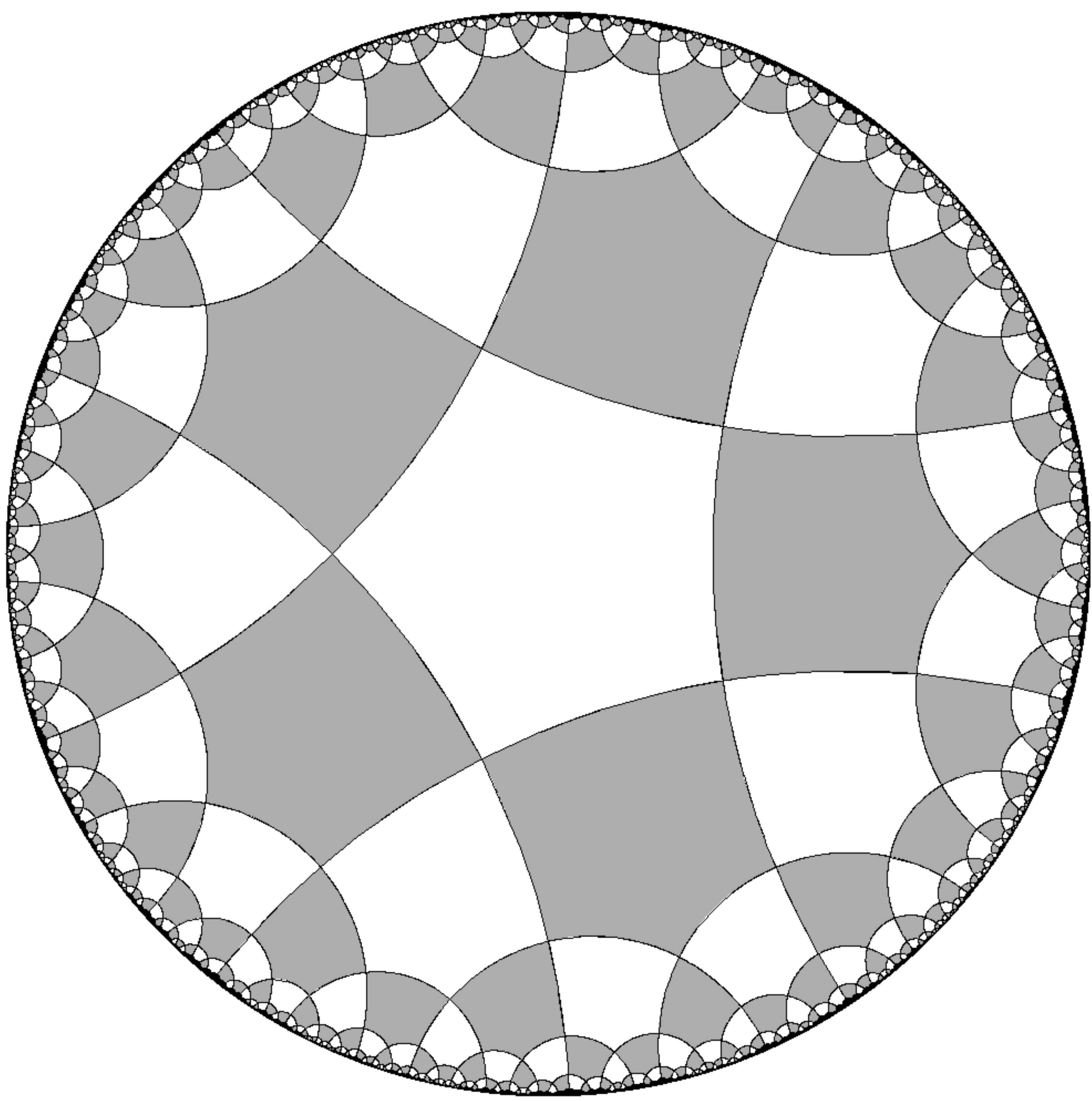}}
\caption{The Euclidean square lattice geometry ($p=4$) and the pentagonal hyperbolic
lattice geometry ($p=5$). Notice that the hyperbolic lattice is made of the regular
pentagons of the equal size and shape. Nevertheless, the mapping of the lattice onto
the Poincare disk depicts them deformed with exponentially decreasing sizes toward the
circle boundary.}
\label{Fig1}
\end{figure}

\begin{figure}[tb]
\centerline{\includegraphics[width=0.4\textwidth,clip]{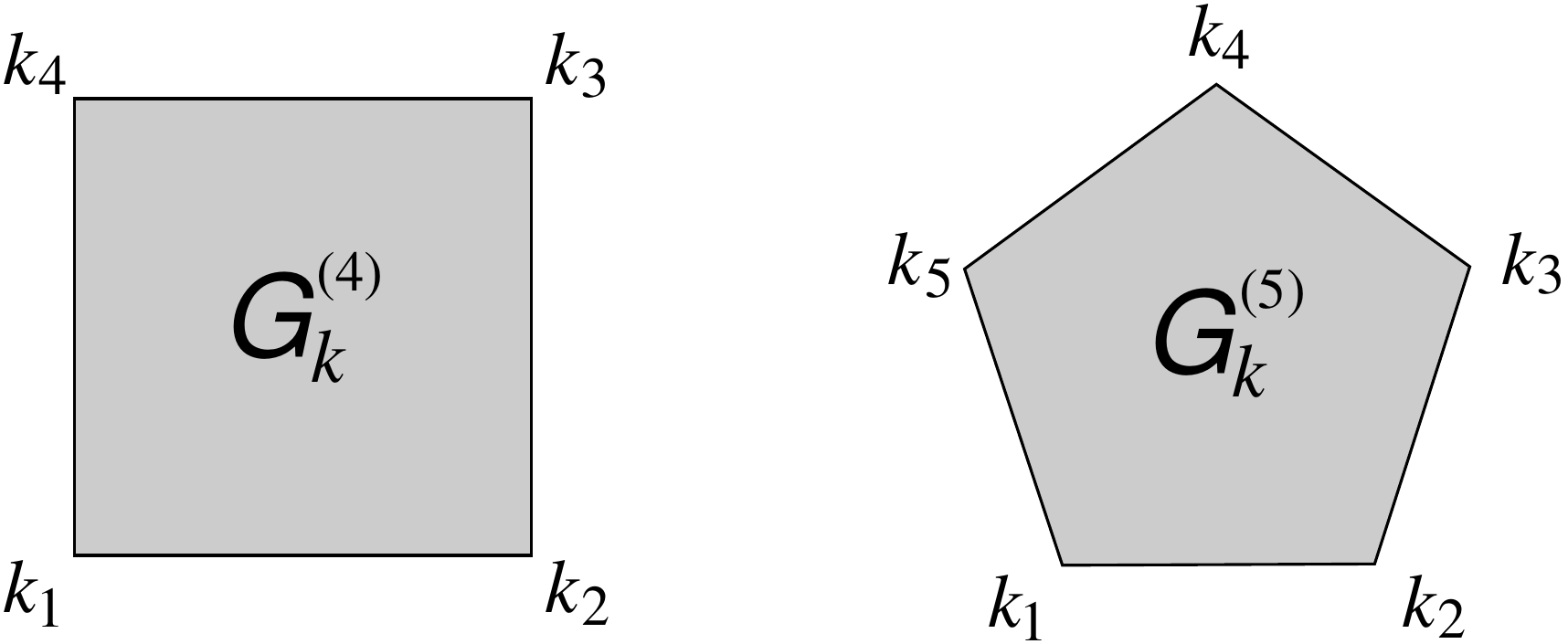}}
\caption{Graphical representation of the local Hamiltonian $G_k^{(p)}$
with its particular shape; the square ($p=4$) on the left and the pentagon ($p=5$)
on the right.}
\label{Fig2}
\end{figure}

We calculate the ground-state energy of the quantum Ising, XY, and Heisenberg models
on two different lattices: the Euclidean square lattice and the hyperbolic pentagonal
lattice. Figure~\ref{Fig1} depicts the two lattice types constructed by tessellation
of the congruent polygons: either by the squares ($p=4$) shown on the left or by the
pentagons ($p=5$) on the right. The alternating gray-white color scheme of the polygons
is chosen for improving the visibility only. The spin variables are located on the
vertices of the regular polygons with the constant coordination number, which is equal
to four in both cases. We study properties of the spin system in the thermodynamic
limit, i.e., the number of the lattice vertices is infinite. The Hamiltonian ${\cal H}$
of the three models defined on the both lattices can be expressed in the following
compact form
\begin{equation}
{\cal H} = \sum\limits_{{\langle k \rangle}_p}^{~} G_{k}^{(p)}\, ,
\label{Hm1}
\end{equation}
where $G_{k}^{(p)}$ represents the local Hamiltonian of the $p$-sided polygon,
the lattice is constructed from, and $k$ marks the position of the polygon
on the lattice. The summation runs over all the positions of the polygons
${\langle k \rangle}_p$. The polygon on the $k^{\rm th}$ position is described
by the ordered set of spin indices $k_1$, $k_2$, ..., $k_p$, see Fig.~\ref{Fig2},
where $k_i$ stands for the unique number which is assigned to the corresponding
vertex within the labeling scheme of the lattice vertices. The local Hamiltonian
has the expression
\begin{equation}
G_{k}^{(p)} = -\frac{1}{2}\sum\limits_{i=1}^{p} \left[
      J_{xy} \left( S_{k_i}^{x} S_{k_{i+1}}^{x} + S_{k_i}^{y} S_{k_{i+1}}^{y} \right)
      + J_{z} S_{k_i}^{z} S_{k_{i+1}}^{z}
      + \frac{h}{4} \left( S_{k_i}^{x} + S_{k_{i+1}}^{x}\right) \right] \, ,
\label{Hm2}
\end{equation}
where $S_{k_i}^x$, $S_{k_i}^y$, $S_{k_i}^z$ are the Pauli operators, and the spin
indices obey the cyclic condition $k_{p+1}\equiv k_1$. The $x$-component of the
external magnetic field is described by the variable $h$, and the spin couplings
$J_{xy}$ and $J_{z}$ specify the three models. In particular, $J_{xy} = -J_{z} = 1$,
and $h=0$ describe the Heisenberg model, $J_{xy} = 1$ and $J_{z} = h = 0$ specify
the XY model, whereas $J_{xy} = 0$, $J_{z} = 1$ with an arbitrary $h$ lead to the
transverse field Ising model~\cite{TPVF,rem}. Since the ferromagnetic ordering of
the models leads to a simpler TPS formulation, we have opted for the positive coupling,
$J_{xy}=1$, and we consider dimensionless units throughout the entire work. The constant
prefactors $\frac{J}{2}$ and $\frac{h}{8}$ reflect the sharing of the spin couplings
and the magnetic field, respectively, if the Hamiltonian is formed by the polygonal
tessellation in Eq.~\eqref{Hm1}.

Our objective is to obtain the ground-state of the system
\begin{equation}
      | \Phi_p\rangle = \lim\limits_{N\to\infty}
      \sum\limits_{\sigma_1^{~}\sigma_2^{~}\cdots\sigma_{N}^{~}}^{~}
      \Phi_p^{\sigma_1^{~}\sigma_2^{~}\cdots\sigma_{N}^{~}}
      |\sigma_1^{~}\sigma_2^{~}\cdots\sigma_N^{~}\rangle
\label{Phi}
\end{equation}
in the thermodynamic limit by a variational minimization of the ground-state energy
normalized per bond
\begin{equation}
{\cal E}_0^{(p)} = \min\limits_{\Phi_p}\lim\limits_{N_b\to\infty}\frac{1}{N_b}
            \frac{\langle \Phi_p | {\cal H} | \Phi_p \rangle}
                 {\langle \Phi_p            | \Phi_p \rangle} \, ,
\label{Eg}
\end{equation}
where $N$ stands for the total number of the lattice spins, $\sigma_j$, $j=1, ..., N$,
marks one of the two base states $\dn$ or $\up$ of the $j^{\rm th}$ lattice spin
and $N_b$ denotes the total number of the bonds (the nearest-neighbor pairs).
In order to simplify the numerical calculation, we approximate $| \Phi_p \rangle$
by a TPS $| \Psi_p \rangle$, which is given by the product of the identical
tensors $W_p$ of the same polygonal structure as either of the two
local Hamiltonians $G_k^{(p)}$ has (cf. Fig.~\ref{Fig2}). The $p$-rank tensors
depend on $p$ spin-$\frac{1}{2}$ variables labeled by indices $k_1, ..., k_p$ with two base states $\sigma_{k_i}^{~}=$ $\dn$ or $\up$.
The $p$ individual spin variables are grouped into a single one with $2^p$ base configurations denoted as $\{\sigma_k\}$ to simplify the notations if necessary.
It means that the tensor element $W_p(\{ \sigma_k \})\equiv W_p(\sigma_{k_1}^{~}\sigma_{k_2}^{~}\cdots
\sigma_{k_p}^{~} )$. For instance, there are 32 base spin configurations for
the pentagons, which can be represented in the arrow notation as $\{\dn\dn\dn\dn\dn\}$,
$\{\dn\dn\dn\dn\up\}$, $\{\dn\dn\dn\up\dn\}$, ..., $\{\up\up\up\up\up\}$.
Thus, the approximative ground state in the form of the polygonal TPS~\cite{Orus,rem2}
has the following form in the thermodynamic limit
\begin{equation}
      | \Psi_p\rangle = \lim\limits_{N\to\infty}
      \sum\limits_{\sigma_1^{~}\sigma_2^{~}\cdots\sigma_{N}^{~}}^{~}
      \prod\limits_{{\langle k \rangle}_p}^{~} W_p(\{\sigma_k\}) 
      |\sigma_1^{~}\sigma_2^{~}\cdots\sigma_N^{~}\rangle \, ,
\label{Psi}
\end{equation}
where the sum runs over
the $2^N$ base spin states. Since the TPS $|\Psi_p\rangle$ has the product structure
of the identical tensors $W_p$, the variational problem in Eq.~\eqref{Eg} is in the
thermodynamic limit equivalent to the minimization of the local energy of an arbitrary
bond in the lattice center (to avoid boundary effects)
\begin{equation}
{E}_0^{(p)} \equiv \min\limits_{\Psi_p}
            \frac{2}{p}\frac{\langle \Psi_p | {G_\ell^{(p)}} | \Psi_p \rangle}
                 {\langle \Psi_p            | \Psi_p \rangle} \, > {\cal E}_0^{(p)} \, ,
\label{Egbond}
\end{equation}
where $\ell$ is the index of a polygon containing the selected central bond and
the factor $2/p$ reflects that each polygon contains $p$ bonds shared with neighboring
polygons. Moreover, the product structure of $|\Psi_p\rangle$ enables us to express the
denominator
\begin{equation}
{\langle \Psi_p | \Psi_p \rangle} = \sum\limits_{\sigma,\sigma^{\prime} }
              \prod\limits_{{\langle k \rangle}_p}^{~}
             W_p^{*}(\{\sigma^{\prime}_k\})  \delta_{\{\sigma^{\prime}_k\},\{\sigma^{~}_k\}}
             W_p(\{\sigma_k\})
            \equiv {\cal D}(W_p(\{\sigma\}))                                         
\label{denominator}
\end{equation} 
and the numerator
\begin{eqnarray}
\nonumber
{\langle \Psi_p | {G_\ell^{(p)}} | \Psi_p \rangle}  &=&  \sum\limits_{\sigma,\sigma^{\prime} }
                         \left[ W_p^{*}(\{\sigma^{\prime}_\ell\})
                                (G_\ell^{(p)})_{\{\sigma^{\prime}_\ell\},\{\sigma^{~}_\ell\}}
                                W_p^{~}(\{\sigma_\ell\}) \right.\\
           &\times &  \prod\limits_{{\langle k \rangle}_p \setminus \{\ell\}}^{~}	
         \left.  W_p^{*}(\{\sigma^{\prime}_k\})
                  \delta_{\{\sigma^{\prime}_k\},\{\sigma^{~}_k\}}
                  W_p^{~}(\{\sigma_k\}) \right] \equiv {\cal N}(W_p(\{\sigma\}))                                
\label{numerator}
\end{eqnarray} 
as sole functions of the tensor elements $W_p(\{\sigma\})$, where we removed the subscript
$k$ due to the uniform TPS. Here, $(G_\ell^{(p)})_{\{\sigma^{\prime}_\ell\},
\{\sigma^{~}_\ell\}}$ stands for the corresponding matrix element of the local Hamiltonian
$G_\ell^{(p)}$, $\delta_{\{\sigma^{\prime}_k\},\{\sigma^{~}_k\}}$ is the Kronecker symbol,
and ${\langle k\rangle}_p \setminus \{\ell\}$ denotes the set of all polygon indices except
for the index $\ell$. 

Consequently, the minimization over the set of variational parameters
$\Phi_p^{\sigma_1^{~}\sigma_2^{~}\cdots\sigma_{\infty}^{~}}$ in Eq.~\eqref{Eg}
is replaced by a much simpler problem
\begin{equation}
{E}_0^{(p)} =  \min\limits_{W_p(\{\sigma\})} 
               \frac{2}{p}\frac{{\cal N}(W_p(\{\sigma\}))}{{\cal D}(W_p(\{\sigma\}))} \, ,
\label{e0}
\end{equation}
where we minimize over $2^p$ tensor elements $W_p(\{\sigma\})$ only. This set can be
further reduced if additional constraints are taken into account. The translational and
rotational symmetries shrink them to three variational parameters, provided that the ground
state $|\Psi_p\rangle$ is evaluated at zero magnetic field, i.e., no symmetry-breaking
mechanism is present in the system. The calculation of the numerator ${\cal N}
(W_p(\{\sigma\}))$ and the denominator ${\cal D}(W_p(\{\sigma\}))$ in Eq.~\eqref{e0}
is carried out separately by means of a numerical algorithm described below.

Thus defined TPS approach offers a new perspective of estimating the 2D quantum
systems on hyperbolic surfaces. Such a significant approximation is sufficient only
for the case of $p=5$ since we have observed a weak entanglement and non-critical
behavior of the classical Ising model on variety of hyperbolic lattices in our
earlier studies~\cite{hctmrg-Ising-5-4,hctmrg-Ising-p-4,hctmrg-Ising-3-q}.
An exponential decay of the density matrix spectra and the correlation function
result in the phase transition, where the correlation length $\xi\lesssim1$ is
finite, reaching its maximal value at the phase transition\cite{corrlen}. For this
reason, we apply the current scheme of TPS as it can be a sufficient approximation
for the ground-state properties of the models on the pentagonal lattice. (However,
the square lattice requires to consider a larger dimension in the tensors $W_4$,
cf.~Refs.\cite{tpvs} and \cite{HoSRG}.)

\subsection{The TPFV algorithm}
   
We refer to the algorithm as the Tensor Product Variational Formulation (TPVF) and
it consists of two parts. The first one evaluates the ratio in Eq.~\eqref{e0}
by applying the CTMRG method~\cite{ctmrg-tn} separately to the numerator and the
denominator for a given set of
the tensor elements $W_p(\{\sigma\})$. The second part contains a multi-dimensional
minimizer, the Nelder-Mead simplex algorithm~\cite{gsl-page,gsl-manual,NelderMead},
which uses the first part to search for the optimized set of the tensor elements
$W_p(\{\sigma\})$, which minimize the ratio in Eq.~\eqref{e0}.
The minimizer starts from an
initial simplex in the space of free variational parameters, one vertex of which is
specified by the initial tensor elements $W_p(\{\sigma\})$. The simplex undergoes an
iterative sequence of size changes and moves towards lower energies and stops if the
energy in Eq.~\eqref{e0} converged. 

\begin{figure}[tb]
\centerline{\includegraphics[width=0.5\textwidth,clip]{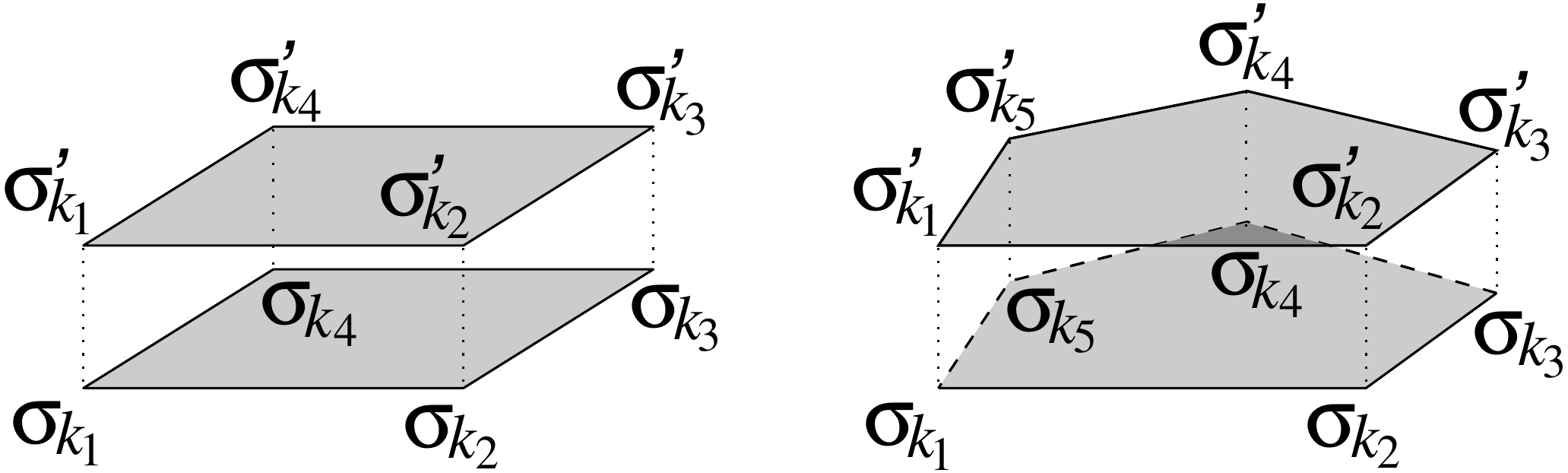}}
\caption{The double-layer tensor structure of ${\cal Z}_4^{~}(\{\sigma_k^\prime
\sigma_k^{~}\})$ on the left and ${\cal Z}_5^{~}(\{\sigma_k^\prime
\sigma_k^{~}\})$ on the right.}
\label{Fig3}
\end{figure}

The CTMRG was originally developed to study 2D classical spin systems. Primarily,
it evaluates the partition function (preferably, in the thermodynamic limit) as the
configuration sum of the tensor product of the Boltzmann weights $W_B$. The central
idea of this study is to apply CTMRG to quantum systems by replacing the concept of
the Boltzmann weight $W_B$ from the classical statistical mechanics by the tensors
$W_p$. In order to do this, let us introduce a double-layer tensor ${\cal Z}_p$
with the tensor elements
\begin{equation}
{\cal Z}_p^{~}(\{\sigma_k^\prime\sigma_k^{~}\}) \equiv
W_p^{*}(\{\sigma^{\prime}_k\}) \delta_{\{\sigma^{\prime}_k\},\{\sigma^{~}_k\}}
W_p^{~}(\{\sigma_k\}) \, .
\end{equation}
Notice that there are $2^{2p}$ double-layer base spin configurations
$\{\sigma_k^\prime\sigma_k^{~}\}$. Figure~\ref{Fig3} graphically depicts the
double-layer tensors ${\cal Z}_p$ at the position $k$, where each shaded polygonal
area represents $W_p$. Thus, in the language of the classical statistical mechanics,
the general expression for the denominator ${\cal D}({W_p (\{\sigma\}}))=\langle\Psi_p
| \Psi_p\rangle$ in Eq.~\eqref{denominator} corresponds to a tensor product object,
which is equivalent to the partition function of a (non-physical) classical Hamiltonian
given by the product of the tensors ${\cal Z}_p$.

This generalization of the Boltzmann weight enters the CTMRG algorithm for both the
lattice geometries and the consequent numerical calculation yields the denominator
${\cal D}(W_p(\{\sigma\}))$ for the given set of the tensor elements $W_p(\{\sigma\})$
according to Eq.~\eqref{denominator}. A similar approach can be also used to determine
the numerator ${\cal N}(W_p(\{\sigma\}))$, as it differs from ${\cal D}(W_p(\{\sigma\}))$
only by the additional double-layer structure at the central position $\ell$ containing
the local Hamiltonian $G_\ell^{(p)}$. The detailed survey of the CTMRG algorithm is
explained in Refs.~\cite{hctmrg-Ising-5-4,hctmrg-Ising-p-4,ctmrg-tn} for both the
square and the pentagonal lattices. We point out only the most relevant part of the
CTMRG algorithm in the following.

\begin{figure}[tb]
\centerline{\includegraphics[width=0.33\textwidth,clip]{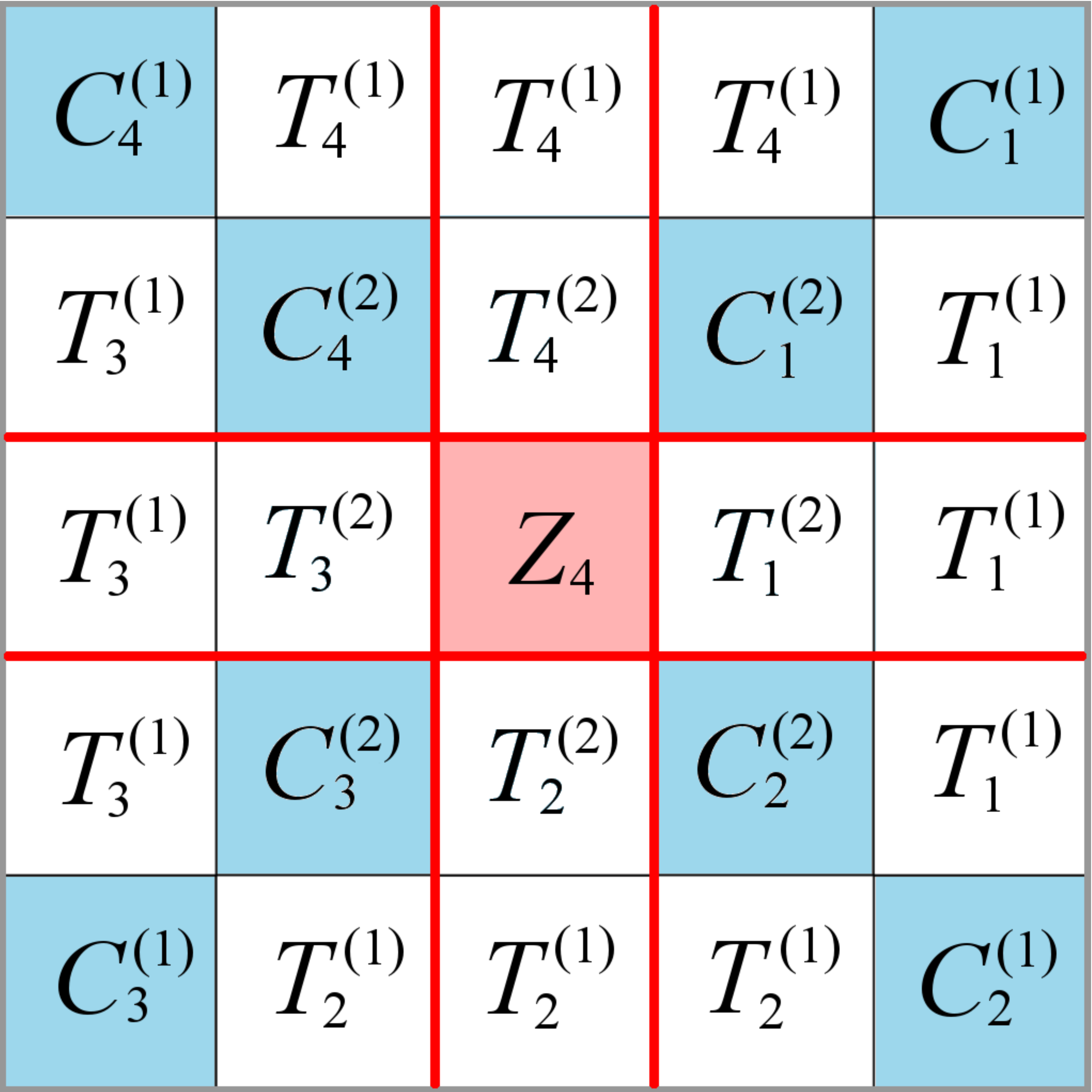}
            \includegraphics[width=0.35\textwidth,clip]{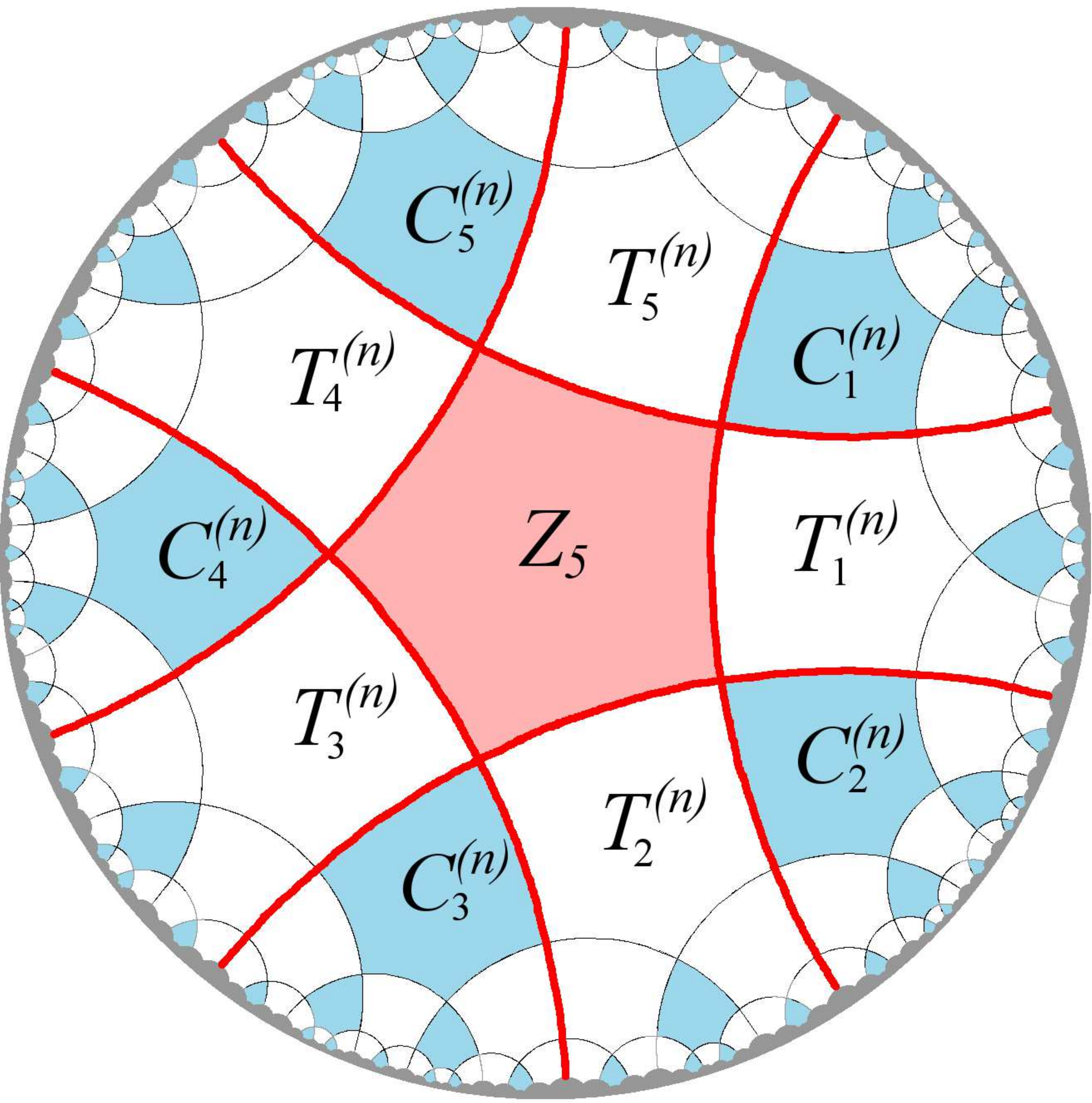}}
\caption{(Color online) The graphical illustration of the denominator in Eq.~\eqref{DWp}
for $p=4$ (left) and $p=5$ (right). The central polygon tensor ${\cal Z}_p$ is surrounded
by $p$ identical (corner) transfer tensors ${\cal C}_j^{(n)}$ and ${\cal T}_j^{(n)}$
bordered by the red thick curves within the $n^{\rm th}$ iteration step in accord with
Eqs.~\eqref{Cn} and \eqref{Tn}. The light blue and the white polygons, respectively,
correspond to ${\cal C}_j^{(n)}$ and ${\cal T}_j^{(n)}$ with the gradually decreasing
iteration step $(n)$ towards the lattice boundary, where $n=1$. Note that the tensors
${\cal C}_j^{(n)}$ and ${\cal T}_j^{(n)}$ are not localized on single polygons only
(as briefly depicted), but all of the tensors for $n=1,2,...,(n-1)$ including
${\cal Z}_p$ are recursively inherited in accord with Eqs.~\eqref{Cn} and \eqref{Tn}.}
\label{Fig4}
\end{figure}

The concept of CTMRG is built on corner transfer tensors ${\cal C}_j^{(n)}$ and transfer
tensors ${\cal T}_j^{(n)}$, where $j=1,2,...,p$. Each of the (corner) transfer tensors 
is composed of the polygon representing tensors ${\cal Z}_p$ via recurrence relations
indexed by the iteration step $n$ (as specified later). Further details of the
construction of the transfer tensors are given in Refs.~\cite{ctmrg-tn,hctmrg-Ising-5-4}.
As a result, the (corner) transfer tensors represent a specific lattice sector formed
by the corresponding polygons. The square and the pentagonal lattices are constructed
from a central polygon surrounded by the alternating sectors represented by the tensors
${\cal C}_j^{(n)}$ and ${\cal T}_j^{(n)}$. The central polygon is represented by the
tensor ${\cal Z}_p$ or $W_p^{*} G_\ell^{(p)}W_p^{~}$ in the structure of the denominator
${\cal D}({W_p (\{\sigma\})})$ and the numerator ${\cal N}({W_p (\{\sigma\})})$,
respectively. Figure~\ref{Fig4} illustrates the situation for the case of
${\cal D}({W_p (\{\sigma\})})$. Consequently, the relations \eqref{denominator} and
\eqref{numerator} for the denominator ${\cal D}({W_p(\{\sigma\})})$ and the numerator
${\cal N}({W_p (\{\sigma\})})$, if formulated in the CTMRG language of the (corner)
transfer tensors, take the form
\begin{equation}
\label{DWp}
{\cal D}({W_p (\{\sigma\})})=
\lim\limits_{n\to\infty} {\rm Tr} \left( {\cal Z}_p
\prod\limits_{j=1}^{p}{\cal C}^{(n)}_j{\cal T}^{(n)}_j \right),
\end{equation}
\begin{equation}
\label{NWp}
{\cal N}({W_p (\{\sigma\})})=
\lim\limits_{n\to\infty} {\rm \Tr} \left( W_p^{*} G_\ell^{(p)}W_p^{~}
\prod\limits_{j=1}^{p}{\cal C}^{(n)}_j{\cal T}^{(n)}_j \right).
\end{equation}

The tensors ${\cal C}_j^{(n)}$ and ${\cal T}_j^{(n)}$ expand their sizes iteratively,
following the recurrence relations \cite{hctmrg-Ising-5-4,hctmrg-Ising-p-4}
\begin{eqnarray}
\label{Cn}
{\cal{C}}_j^{(n)}={\cal Z}_p{\cal{T}}_{j-1}^{(n-1)}
 \left[{\cal{C}}_j^{(n-1)}{\cal{T}}_j^{(n-1)}\right]^{p-3} \, , \\
\label{Tn}
{\cal{T}}_j^{(n)}={\cal Z}_p {\cal{T}}_{j-1}^{(n-1)}
 \left[{\cal{C}}_j^{(n-1)}{\cal{T}}_j^{(n-1)}\right]^{p-4} \, ,
\end{eqnarray}
where the transfer tensors satisfy the cyclic condition ${\cal T}_0^{(n)}\equiv
{\cal T}_p^{(n)}$. The infinite TPS geometry is built up gradually by increasing
the iteration step $n=2,...\infty$, which induces an exponential increase of the
degrees of freedom of the (corner) transfer tensors. The calculations are kept
numerically feasible by means of the renormalization group procedure, which
integrates out the least probable configurations in the tensors determined by
the reduced density matrix~\cite{dmrg-srw,ctmrg-tn}. 

The recurrence relations are initialized in the first iteration step, $n=1$ by
construction of the tensors ${\cal C}_j^{(n=1)}$ and ${\cal T}_j^{(n=1)}$ from the
tensor ${\cal Z}_p$ for all $j=1,2,...,p$. If the spin variables $\sigma_{k_i}$ are
explicitly included in the tensors, we have
\begin{eqnarray}
\nonumber
{\cal C}_j^{(1)}(\{\sigma_{k_1}^\prime\sigma_{k_2}^\prime\sigma_{k_3}^\prime
                  \sigma_{k_1}^{~}   \sigma_{k_2}^{~}   \sigma_{k_3}^{~}\})
&=& \sum\limits_{\genfrac{}{}{0pt}{}
                   {\sigma_{k_5}^\prime\sigma_{k_4}^\prime}
                         {\sigma_{k_5}^{~}   \sigma_{k_4}^{~}}}
    {\cal Z}_5(\{\sigma_k^\prime\sigma_k^{~}\}) \, , \\ \nonumber
&=& \sum\limits_{\sigma_{k_4}^\prime\sigma_{k_4}^{~}}
    {\cal Z}_4(\{\sigma_k^\prime\sigma_k^{~}\}) \, , \\ \nonumber
{\cal T}_j^{(1)}
(\{\sigma_{k_1}^\prime \sigma_{k_2}^\prime \sigma_{k_3}^\prime \sigma_{k_4}^\prime
  \sigma_{k_1}^{~}    \sigma_{k_2}^{~}    \sigma_{k_3}^{~}    \sigma_{k_4}^{~}\})
&=& \sum\limits_{\sigma_{k_5}^\prime \sigma_{k_5}^{~}}
    {\cal Z}_5(\{\sigma_k^\prime \sigma_k^{~}\}) \, , \\
&=& {\cal Z}_4(\{\sigma_k^\prime \sigma_k^{~}\}) \, .
\end{eqnarray}

\subsection{The tensor symmetries} \label{freeparam}

As mentioned earlier, let the polygon base spin configuration $\{\sigma_k^{~}\}\equiv
(\sigma_{k_1}^{~}\sigma_{k_2}^{~}\cdots\sigma_{k_p}^{~})$ be
given in the spin-arrow notation. Let function $\theta_j$ return all such configurations
for which the integer number $j$ counts the number of the spins aligned upward as
listed in Tabs.~\ref{Tab1} and \ref{Tab2}. Each line in the Tables contains such
spin configurations, which are identical with respect to the rotational
symmetry operations of the $p$-sided polygon. We make a difference
between the spin configurations $\theta_j^{~}$ and $\theta_{j^\prime}^{~}$ when $j=2$
(or $j=3$ for the case of $p=5$) as they are not rotationally equivalent. The prime
symbol is used for such configuration (and its rotations), in which either the two
($j=2$) or three ($j=3$) spins, aligned upward in each polygon, cannot be grouped
together (or equivalently, if the alternating spin alignment is maximal).

\begin{table}[!b]
\caption{The full spin configurations for $p=4$ sorted by $\theta_j$. The last
column lists the reduced parameter set $a_{\ell}^{~}$ with $\ell=\min\{j,p-j\}$.}
\begin{center}
\begin{tabular}{|l|l|cccc|c|}
\hline
 
$\, j$ & $W_4^{~}(\theta_j^{~})$ & \multicolumn{4}{ c |} {$\theta_j^{~} \equiv \{\sigma_{k_1}^{~}\sigma_{k_2}^{~}\sigma_{k_3}^{~}\sigma_{k_4}^{~}\}$} & $a_{\ell}^{~}$ \\[0.1cm]
\hline

 $0$        & $W_4^{~}(\theta_0^{~})$ & $\{\dn\dn\dn\dn\}$ &                  &                  &                  & $a_0^{~}$ \\
 $1$        & $W_4^{~}(\theta_1^{~})$ & $\{\dn\dn\dn\up\}$ & $\{\dn\dn\up\dn\}$ & $\{\dn\up\dn\dn\}$ & $\{\up\dn\dn\dn\}$ & $a_1^{~}$ \\
 $2$        & $W_4^{~}(\theta_2^{~})$ & $\{\dn\dn\up\up\}$ & $\{\dn\up\up\dn\}$ & $\{\up\up\dn\dn\}$ & $\{\up\dn\dn\up\}$ & $a_2^{~}$ \\
 $2^\prime$ & $W_4^{~}(\theta_{2^\prime}^{~})$ & $\{\dn\up\dn\up\}$ & $\{\up\dn\up\dn\}$ &                  &         & \ $a_{2^\prime}$ \\
 $3$        & $W_4^{~}(\theta_3^{~})$ & $\{\up\up\up\dn\}$ & $\{\up\up\dn\up\}$ & $\{\up\dn\up\up\}$ & $\{\dn\up\up\up\}$ & $a_1^{~}$ \\
 $4$        & $W_4^{~}(\theta_4^{~})$ & $\{\up\up\up\up\}$ &                  &                  &                  & $a_0^{~}$ \\[0.1cm]
\hline
\end{tabular}
\end{center}
\label{Tab1}
\end{table}

\begin{table}[!bt]
\caption{The spin configurations for $p=5$ as in Tab~\ref{Tab1}.}
\begin{center}
\begin{tabular}{|l|l|ccccc|c|}
\hline
 $\, j$ & $W_5^{~}(\theta_j^{~})$ & \multicolumn{5}{ c |} {$\theta_j^{~}\equiv\{\sigma_{k_1}^{~}\sigma_{k_2}^{~}\sigma_{k_3}^{~}\sigma_{k_4}^{~}\sigma_{k_5}^{~}\}$} & $a_{\ell}^{~}$ \\[0.1cm]
\hline
 $0$        & $W_5^{~}(\theta_0^{~})$ & $\{\dn\dn\dn\dn\dn\}$ &                     &                     &                     &                     & $a_0^{~}$ \\
 $1$        & $W_5^{~}(\theta_1^{~})$ & $\{\dn\dn\dn\dn\up\}$ & $\{\dn\dn\dn\up\dn\}$ & $\{\dn\dn\up\dn\dn\}$ & $\{\dn\up\dn\dn\dn\}$ & $\{\up\dn\dn\dn\dn\}$ & $a_1^{~}$ \\
 $2$        & $W_5^{~}(\theta_2^{~})$ & $\{\dn\dn\dn\up\up\}$ & $\{\dn\dn\up\up\dn\}$ & $\{\dn\up\up\dn\dn\}$ & $\{\up\up\dn\dn\dn\}$ & $\{\up\dn\dn\dn\up\}$ & $a_2^{~}$ \\
 $2^\prime$ & $W_5^{~}(\theta_{2^\prime}^{~})$ & $\{\dn\dn\up\dn\up\}$ & $\{\dn\up\dn\dn\up\}$ & $\{\dn\up\dn\up\dn\}$ & $\{\up\dn\dn\up\dn\}$ & $\{\up\dn\up\dn\dn\}$ & \ $a_{2^\prime}$ \\
 $3$        & $W_5^{~}(\theta_3^{~})$ & $\{\up\up\up\dn\dn\}$ & $\{\up\up\dn\dn\up\}$ & $\{\up\dn\dn\up\up\}$ & $\{\dn\dn\up\up\up\}$ & $\{\dn\up\up\up\dn\}$ & $a_2^{~}$ \\
 $3^\prime$ & $W_5^{~}(\theta_{3^\prime}^{~})$ & $\{\up\up\dn\up\dn\}$ & $\{\up\dn\up\up\dn\}$ & $\{\up\dn\up\dn\up\}$ & $\{\dn\up\up\dn\up\}$ & $\{\dn\up\dn\up\up\}$ & \ $a_{2^\prime}$ \\
 $4$        & $W_5^{~}(\theta_4^{~})$ & $\{\up\up\up\up\dn\}$ & $\{\up\up\up\dn\up\}$ & $\{\up\up\dn\up\up\}$ & $\{\up\dn\up\up\up\}$ & $\{\dn\up\up\up\up\}$ & $a_1^{~}$ \\
 $5$        & $W_5^{~}(\theta_5^{~})$ & $\{\up\up\up\up\up\}$ &                     &                     &                     &                     & $a_0^{~}$ \\[0.1cm]
\hline
\end{tabular}
\end{center}
\label{Tab2}
\end{table}

As a consequence, we count six or eight distinguishable configurations $\theta_0\cdots
\theta_p$ for $p=4$ and $p=5$, respectively. These configurations determine the number
of the variational parameters $W_p(\theta_j)$ used in the calculation of the transverse field Ising model. 
If, however, the spontaneous symmetry-breaking does not affect the solution, the
total number of the variational parameters decreases down to four parameters for
the Hamiltonian models (for both lattices $p$). Therefore, the spin-inversion symmetry 
reduces the variational parameters due to the absence of the preferred 
direction of the spontaneous magnetization. In particular, the spin configuration
probability for the pairs $\theta_0$ and $\theta_p$, $\theta_1$ and $\theta_{p-1}$, etc. becomes identical, which
can be formally generalized into the equations
\begin{equation}
\begin{tabular}{l}
$W_p (\theta _j        ) = W_p (\theta_{p- j        }) \equiv a_{\min\{j,p-j\}}^{~}$,\\[0.3cm]
$W_p (\theta_{j^\prime}) = W_p (\theta_{(p-j)^\prime}) \equiv 1$.
\end{tabular}
\label{coupled}
\end{equation}
There are the three equations in the upper expression defining the new variational parameters
$a_0^{~}$, $a_1^{~}$, $a_2^{~}$, and one equation in the lower expression for
$a_{2^{\prime}}^{~}$, which has already been eliminated from the set of the free parameters
by putting $a_{2^{\prime}}^{~}\equiv1$ being the normalization condition in $W_p$.
Hence, the three free variational parameters suffice to approximate the ground-state wave
function of the models with no spontaneous symmetry-breaking phases. For the same reason,
if we consider the system without the spin-inversion symmetry, there are either five or
seven free variational parameters for $p=4$ or $p=5$, respectively, excluding the normalization parameter $W_p(\theta_{2^{\prime}}^{~}) \equiv 1$. 

The number of the operations performed by the implementation of the Nelder-Mead algorithm
we used scales linearly with the number of the free variational parameters~\cite{gsl-manual}.
Of course, increasing the number of the parameters prolongs the computational time,
and may encounter numerical instability caused by trapping the system in a local
minimum of the energy, rather than approaching to the correct global minimum, which
corresponds to $E_0^{(p)}$.
On the other hand, a faster Nelder-Mead optimization enables to improve the
accuracy by increasing the number of the CTMRG states kept (the effective block spin
states~\cite{ctmrg-tn,dmrg-srw}) and make TPVF more efficient.

\section{Numerical results}

As we have mentioned earlier, we consider two types of the lattices. The Euclidean
one is defined on the regular square lattice characterized by $p=4$, which serves as
a reference for the hyperbolic pentagonal lattice with $p=5$ (cf. Fig.~\ref{Fig1}),
which is of our main interest.
First, we start analyzing the ground-state properties of the transverse field Ising
model (TFIM) in details, where the number of the variational parameters
$W_p^{~}(\theta_j^{~})$ is either five (for $p=4$) or seven (for $p=5$), respectively,
excluding the normalization parameter $W_p(\theta_{2^{\prime}}^{~}) \equiv 1$. We analyze the TFIM and its phase transition by the
following three ways: (1) by evaluating the optimized parameters $W_p^{~}(\theta_j^{~})$, (2) by the
expectation value of the magnetization constructed from the ground state $|\Psi_p\rangle$,
and (3) by the ground-state energy per bond $E_0^{(p)}$ and its second derivative.
Second, we reduce the number of the free variational parameters down to three (if all
symmetries are considered) and calculate the ground-state energies of the XY and
Heisenberg models.

In our numerical calculations we keep at most 20 effective block spin
states~\cite{ctmrg-tn}, which is sufficient to reach more than six-digit convergence
in ${\cal N}({W_p (\{\sigma\})})$ and ${\cal D}({W_p (\{\sigma\})})$. Notice that the
numerical accuracy is mainly given by the uniform TPS approximation. Further
improvement of the numerical accuracy requires to consider a non-uniform TPS
(i.e. violation of the translational lattice symmetry), and a gradual
expansion of the tensor order by implementation of auxiliary variables~\cite{tpvs}
is necessary. This is, however, beyond the scope of our interest. 

The Nelder-Mead algorithm has been tested for various initial conditions, most of them
leading to the identical results. Typically, we started with a simplex, where we set all
the coordinates of one of its vertices to unity (the vertex coordinates represent the
free variational parameters $W_p^{~}(\theta_j^{~})$). The initial simplex size was set to
$0.1$~\cite{gsl-page,gsl-manual}.

\begin{figure}[tb]
\centerline{\includegraphics[width=0.7\textwidth,clip]{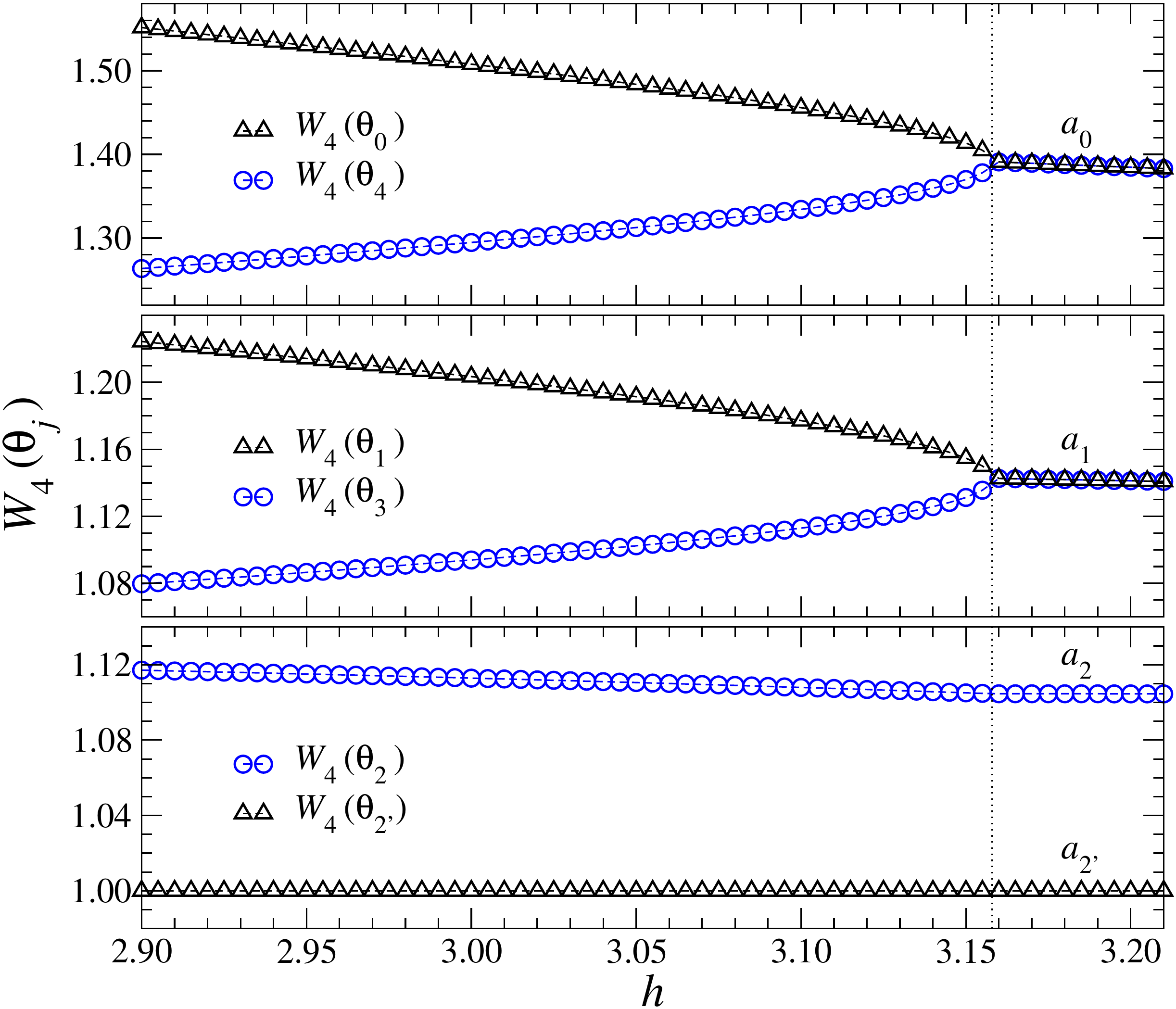}}
\caption{(Color online) The magnetic field dependence of the six variational parameters
$W_4^{~}(\theta_j^{~})$ on the reference square lattice ($p=4$). The singular behavior
corresponds to the phase transition critical field $h_c^{(4)}\approx3.158$.}
\label{Fig5}
\end{figure}

\begin{figure}[tb]
\centerline{\includegraphics[width=0.7\textwidth,clip]{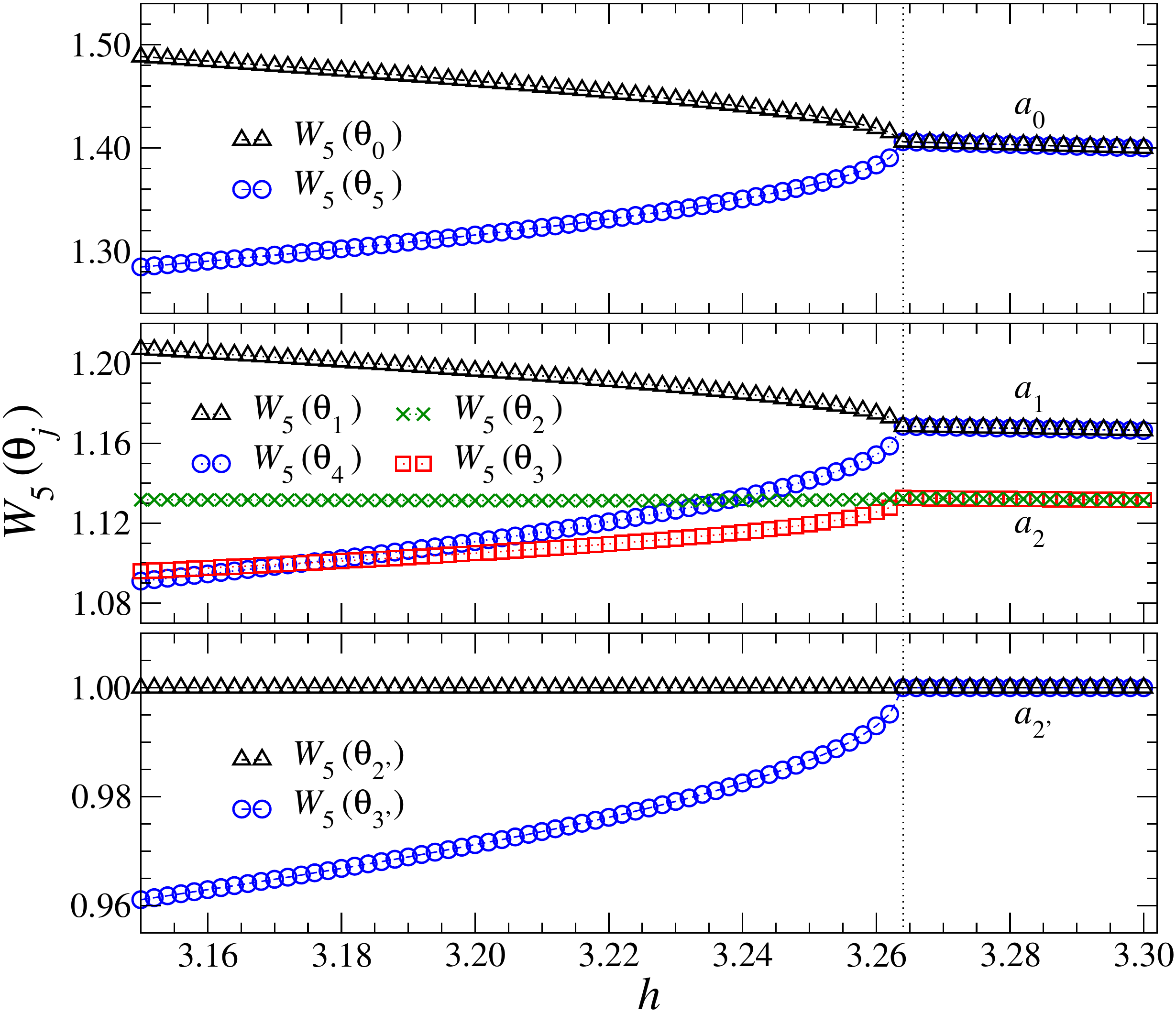}}
\caption{(Color online) The magnetic field dependence of the eight variational parameters
$W_5^{~}(\theta_j^{~})$ on the hyperbolic pentagonal lattice ($p=5$) with the phase transition
field $h_c^{(5)}\approx3.264$.}
\label{Fig6}
\end{figure}

Figures~\ref{Fig5} and \ref{Fig6} illustrate behavior of the optimized variational
parameters $W_p^{~}(\theta_j^{~})$ in the TFIM with respect to the external magnetic field
$h$ for the particular lattice geometry $p$. It is evident that above a certain magnetic
field (depicted by the vertical dotted lines), specific pairs of the variational
parameters collapse onto identical values, which yields four single curves in this region for both the lattices. These pairs are exactly those coupled by the
system of equations in Eq.~\eqref{coupled}, i.e. those representing spin configurations
which are equivalent with respect to spin inversion. The magnetic field, at which the
collapse causes a singular behavior of $W_p^{~}(\theta_j^{~})$, corresponds to the
quantum phase transition of TFIM at the magnetic field $h_c^{(p)}$, which we analyze
below. In the ordered phase at $h<h_c^{(p)}$, the distinct optimized values of the
coupled parameters $W_p^{~}(\theta_j^{~})$, as specified in Tabs.~\ref{Tab1} and
\ref{Tab2}, reflect the existence of the spontaneous symmetry-breaking
in the TFIM for both the lattice types. In the disordered phase at $h\geq h_c^{(p)}$,
the four-parameter description coincides with the free variational parameters $a_0^{~}$,
$a_1^{~}$, $a_2^{~}$, and the normalization parameter $a_{2^{\prime}}^{~}$.
This confirms the relevance of the
additional symmetries in such systems, where the spontaneous symmetry-breaking mechanism
is not present, such as in the XY and Heisenberg systems at the zero magnetic field.

\begin{figure}[tb]
\centerline{\includegraphics[width=0.7\textwidth,clip]{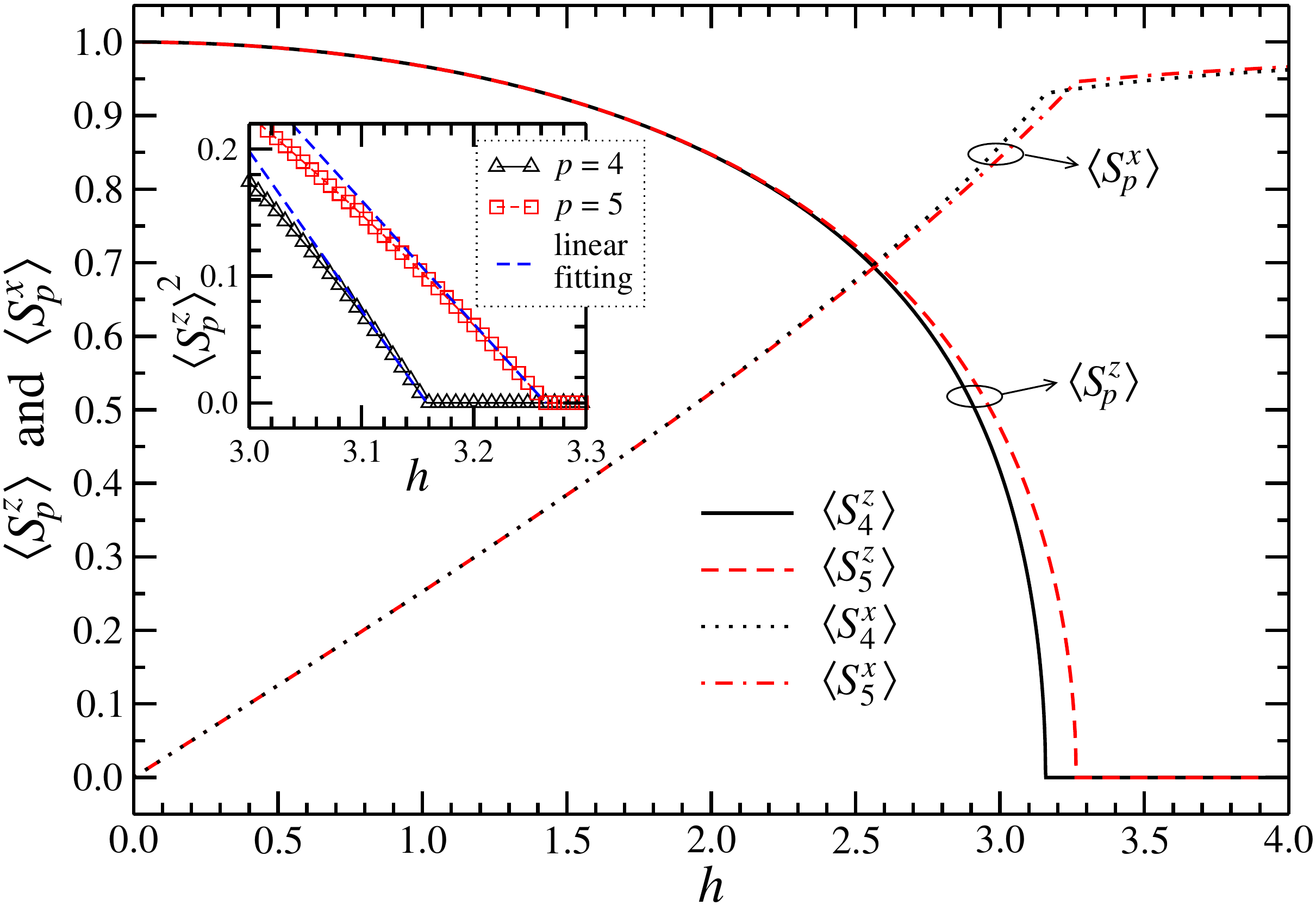}}
\caption{(Color online) The transversal $\langle S_p^{z}\rangle$ and the parallel
$\langle S_p^{x}\rangle$ magnetizations with respect to the magnetic field for
both lattice geometries. The inset shows the mean-field dependence of the magnetization
when approaching the transition field; the linearity is depicted by the blue dashed lines.}
\label{Fig7}
\end{figure}

Having evaluated the optimized free variational parameters, we can easily reconstruct
the approximative eigenstate $|\Psi_p\rangle$ as the TPS and apply it for the evaluation
of the spontaneous magnetization
\begin{equation}
\langle S_p^{\alpha} \rangle = \frac{\langle \Psi_p | S_{k_i}^{\alpha} |\Psi_p \rangle}
                                    {\langle \Psi_p |                   \Psi_p \rangle} \, ,
\label{sa}
\end{equation}
where $\alpha=x$ or $z$. The expectation value of the spin operator $S_{k_i}^{\alpha}$
is evaluated in the central part of the lattice in order to suppress all boundary effects.
Here, $\langle S_p^{z}\rangle$ denotes the order parameter of TFIM and specifies the
quantum phase transition at the phase transition field. The resulting dependence of 
magnetization $\langle S_p^{z}\rangle$ and $\langle S_p^{x}\rangle$ with respect
to the magnetic field $h$ is shown in Fig.~\ref{Fig7}. The quantum phase transition is
reflected by the singular behavior of all magnetization profiles.

Analyzing the TFIM, we calculated $h_c^{(4)}=3.158$ for the Euclidean square geometry and
$h_c^{(5)}=3.264$ for the hyperbolic pentagonal lattice. The most relevant value of the
critical magnetic field for the TFIM on the 2D Euclidean lattice (as obtained by a recent
Tensor RG algorithm) yields $h_c^{\rm TRG} = 3.0439$~\cite{HoSRG}. Since the TPS is built
up by the tensors $W_4$ of the too low dimension, the long-range order correlations are
excluded~\cite{hctmrg-Ising-3-q}, and the mean-field-like dependence near the phase transition
is necessarily observed. The expectation value of the magnetization $\langle S_p^{z}\rangle$
obeys the scaling relation
\begin{equation}
\langle S_p^z (h) \rangle \propto {\left(h_c^{(p)} - h \right)}^{\beta_p} \, .
\label{powerlaw}
\end{equation}
The mean-field exponent $\beta_p=\frac{1}{2}$ is observed
regardless of the lattice geometry. As a reference, the numerical TRG analysis gives
correct $\beta_4^{\rm TRG} = 0.3295$ on the square lattice (free of any mean-field
approximations), which is also in agreement with Monte Carlo simulations~\cite{HoSRG}.
The inset of Fig.~\ref{Fig7} displays the magnetization squared, which confirms the
mean-field-like behavior for both $p$ by its linear dependence on $h$ if approaching the
phase transition. The blue dashed lines serve as guides for the eye in order to enhance
the linearity.

\begin{figure}[tb]
\centerline{\includegraphics[width=0.7\textwidth,clip]{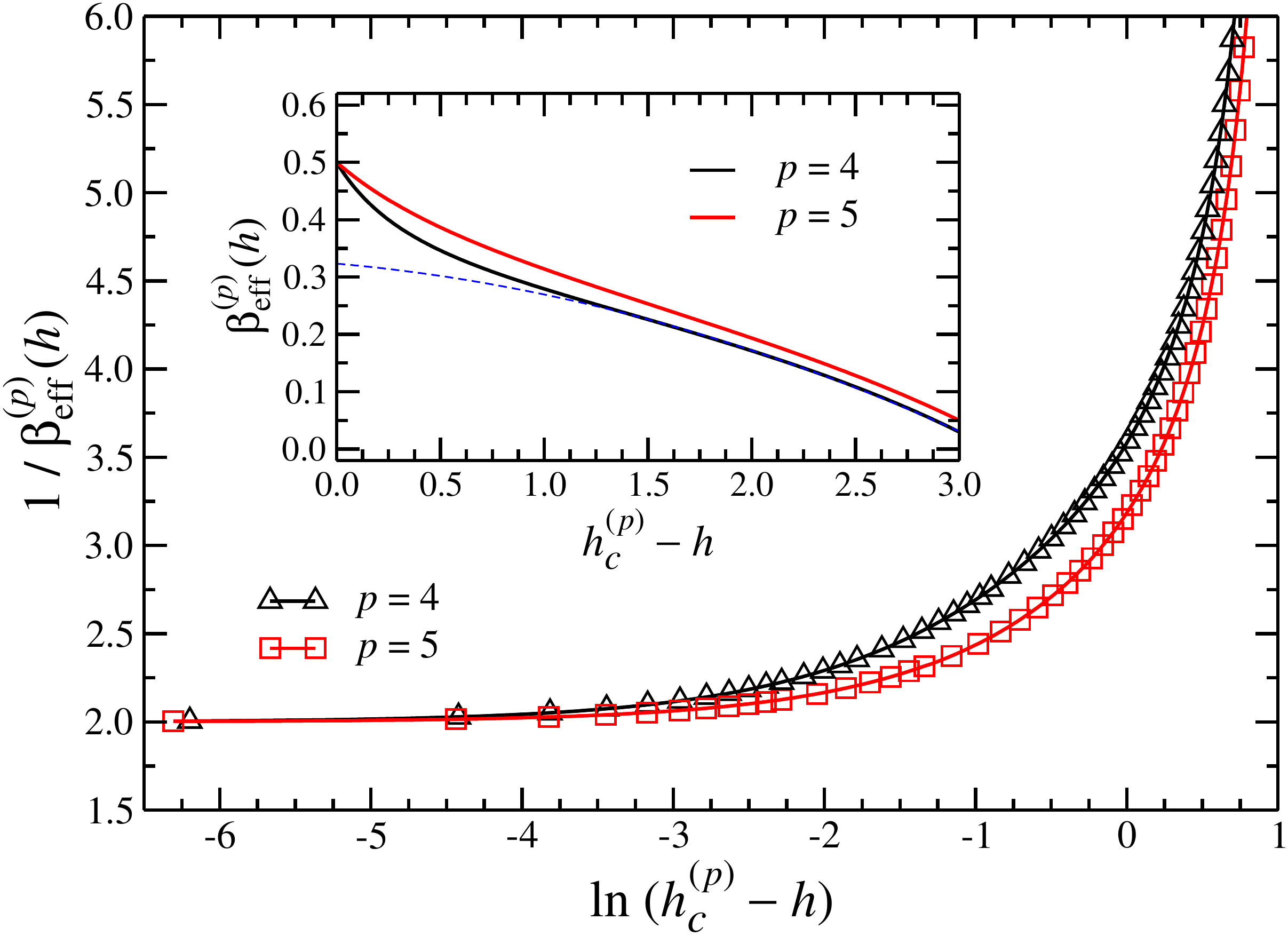}}
\caption{(Color online) The detailed dependence of the inverse effective magnetic
exponent on the magnetic field in the logarithmic form. The inset shows the effective
exponent at wider magnetic field scale. The blue dashed line estimates behavior of
the correct effective exponent for the Euclidean lattice.}
\label{Fig8}
\end{figure}

A more detailed analysis of the TPVF approximation near the phase transition can be
visualized by evaluating the effective (field dependent) exponent $\beta^{(p)}_{\rm eff}(h)$,
which converges to $\beta_p$ when approaching the phase transition field $h_c^{(p)}$
\begin{equation}
\beta_p = \lim\limits_{h \to h_c^{(p)}} \beta^{(p)}_{\rm eff} (h)
        = \lim\limits_{h \to h_c^{(p)}} \frac{\partial \ln\langle S_p^z (h) \rangle}
                                             {\partial \ln \left(h_c^{(p)} -  h \right)}
\label{beta_eff}
\end{equation}
as plotted in Fig.~\ref{Fig8}. The effective exponent obviously converges to the
mean-field exponent $\beta_p=\frac{1}{2}$ for both lattice types if the phase transition
field is approached from the ordered phase, i.e., if $\ln(h_c^{(p)}-h) \to-\infty$. The
inset shows the same dependence on larger scales. The critical exponent on the square
lattice (the black curve for $p=4$) starts deviating at around $h>2.0$ from the expected
exponent (estimated by the blue dashed curve), which is known to converge to
$\beta_4^{\rm TRG}=0.3295$~\cite{HoSRG}.

\begin{figure}[tb]
\centerline{\includegraphics[width=0.7\textwidth,clip]{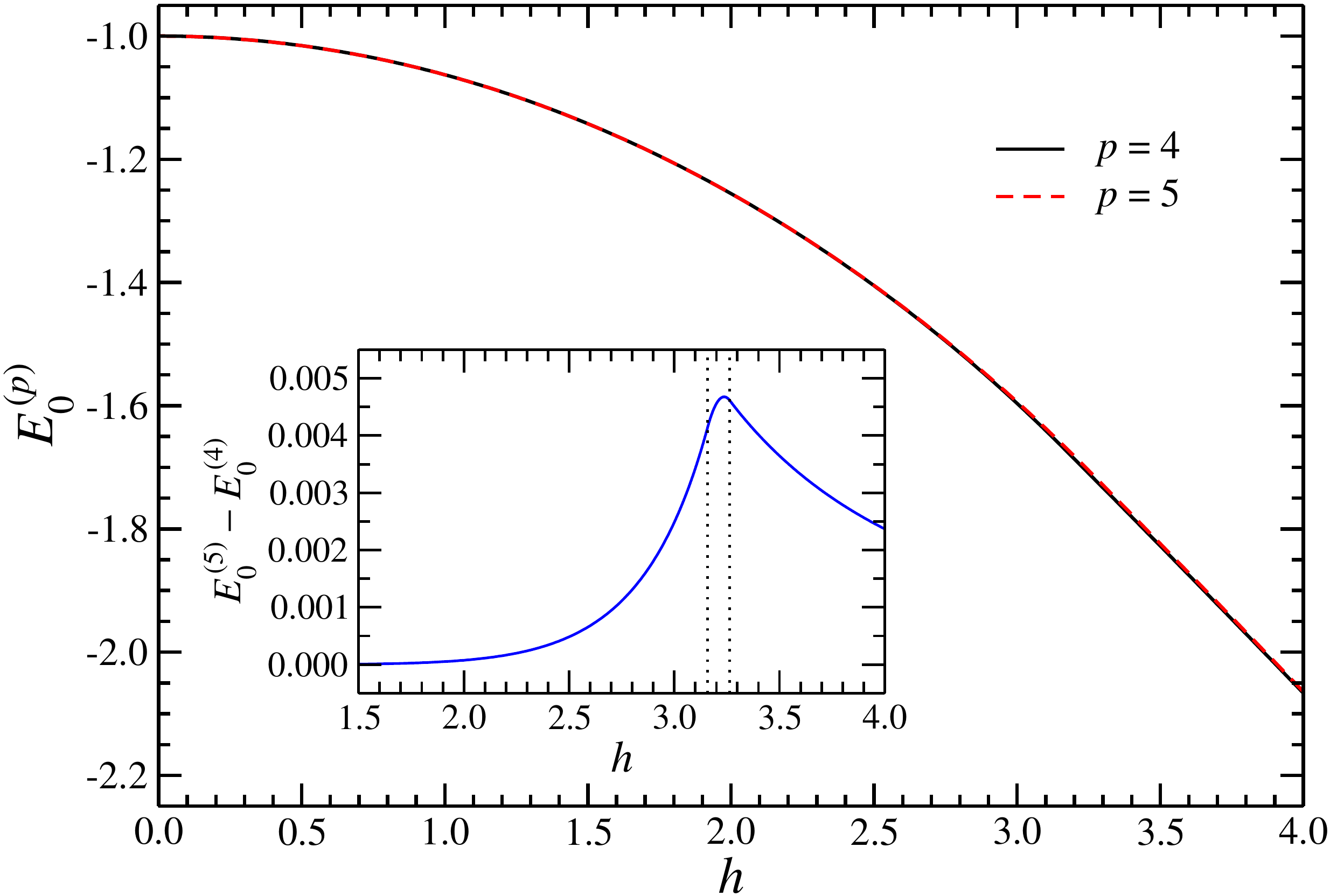}}
\caption{(Color online) The ground-state energy of the TFIM with respect to the
magnetic field $h$. The small difference between the two energies is plotted in the
inset.}
\label{Fig9}
\end{figure}

\begin{figure}[tb]
\centerline{\includegraphics[width=0.7\textwidth,clip]{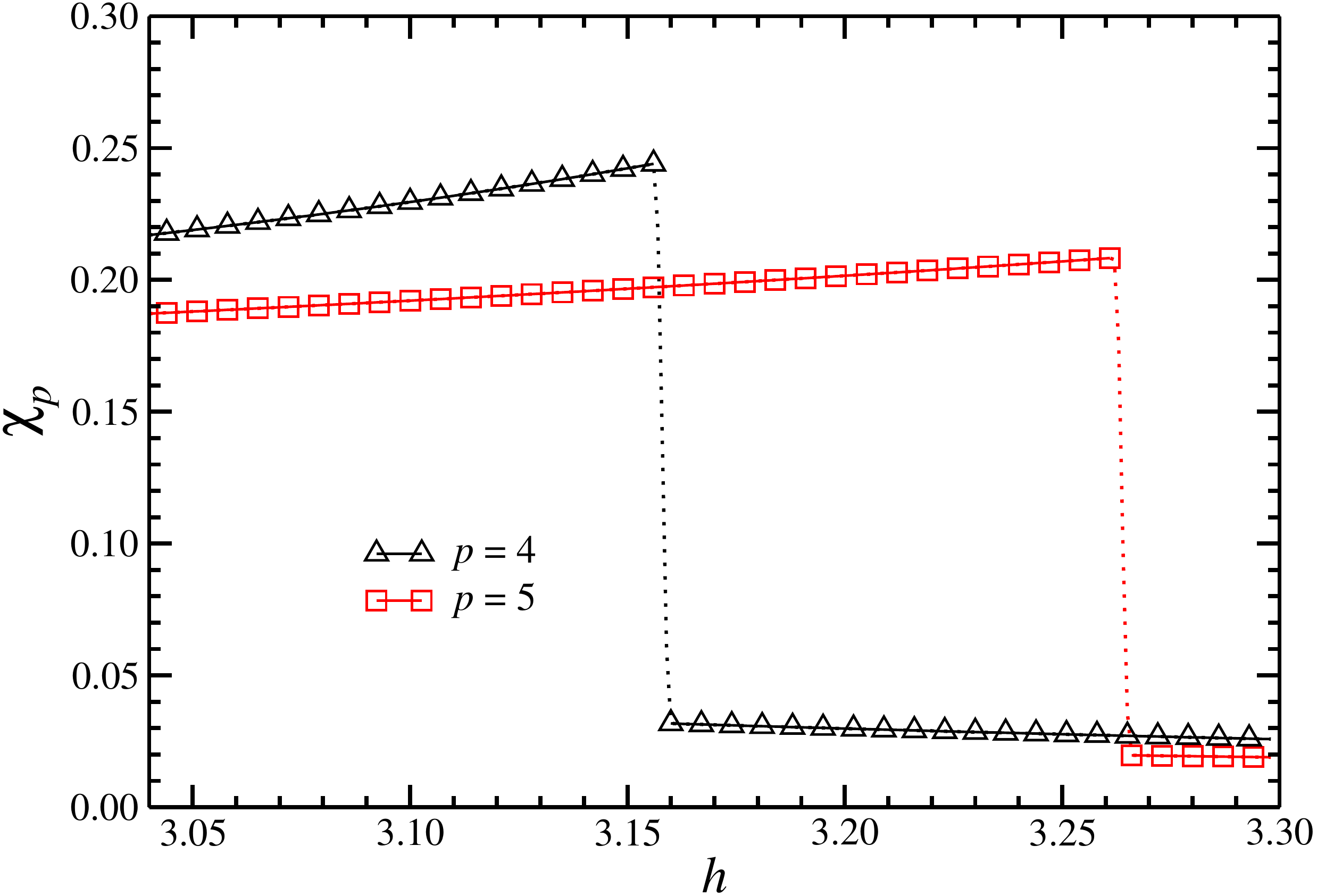}}
\caption{(Color online) The magnetic susceptibility $\chi_p$ of the TFIM
as a function of $h$ for both lattices.}
\label{Fig10}
\end{figure}

The ground-state energies normalized per bond $E_0^{(p)}$ given by Eq.~\eqref{e0} are depicted
in Fig.~\ref{Fig9}. There is a tiny difference in both of the energies. Therefore, we
plotted the energy difference $E_0^{(5)}-E_0^{(4)}$ in the inset (which has no physical
meaning) in order to point out the difference. The two vertical lines in the inset
correspond to the phase transitions (for reference only). The magnetic susceptibility
\begin{equation}
\chi_p=-\frac{\partial^2 E_0^{(p)}}{\partial h^2}
\label{chi}
\end{equation}
is an independent physical quantity, which characterizes the phase transition.
Figure~\ref{Fig10} shows the functional dependence of the susceptibility on the
magnetic field. The calculation of $\chi_p$ requires a precise data for
the ground-state energy because of performing the second derivative numerically.
The shape of the non-diverging discontinuity in $\chi_p$ at the phase transition
magnetic field $h_c^{(p)}$ is the typical consequence of the mean-field universality.

The XY and the Heisenberg models are critical (at $p=4$) at zero magnetic field.
Therefore, the full tensor symmetries are considered, cf. Tabs.~\ref{Tab1} and \ref{Tab2},
resulting in the three free variational parameters $a_0^{~}$, $a_1^{~}$, and
$a_2^{~}$, which is sufficient to obtain the ground-state energy per bond
$E_0^{(p)}$. Table~\ref{Tab3} summarizes thus calculated optimized free
variational tensor parameters and the energies. The comparison of $E_0^{(4)}$ on
the square lattice with the Monte Carlo results by Sandvik~\cite{XY,Heis} serves
as an estimate of the exact ground-state energy ${\cal E}_0^{(5)}$ for the XY
and the Heisenberg models on the hyperbolic pentagonal lattice, which has not been
determined yet.

\begin{table}[ht]
\caption{List of the optimized free variational parameters $a_j^{~}$ which minimize
the ground-state energy per bond $E_0^{(p)}$ on the square and pentagonal lattices for
the XY and Heisenberg models. The ground-state energies $E_0^{\rm MC}$ on the square
lattice are obtained by Monte Carlo simulations~\cite{XY,Heis}.}
\begin{center}
\begin{tabular}{| c | c | c | c | c |}
\hline
 lattice & \multicolumn{2}{ c |}{$p=4$} & \multicolumn{2}{ c |}{$p=5$} \\ \cline{2-5}
 model & XY & Heisenberg & XY & Heisenberg \\ \hline
$a_0^{~}$      & $0.69333$ & $0.50746$ & $0.66001$ & $0.48553$ \\ \hline
$a_1^{~}$      & $0.88010$ & $0.74826$ & $0.83844$ & $0.71804$ \\ \hline
$a_2^{~}$      & $0.94236$ & $0.83422$ & $0.90733$ & $0.80263$ \\ \hline
$E_0^{(p)}$      & $-1.0846${\phantom 0} & $-1.3089${\phantom 0} &
                 $-1.0815${\phantom 0} & $-1.2913${\phantom 0} \\ \hline
$E_0^{\rm MC}$ & $-1.09765$ & $-1.33887$ & --- & --- \\ \hline
\end{tabular}
\end{center}
\label{Tab3}
\end{table}

\section{Conclusions}

We have applied the TPVF algorithm to the calculation of the quantum systems
(transverse field Ising, XY, and Heisenberg models) on the pentagonal hyperbolic
lattice with the constant negative Gaussian curvature and the coordination number
fixed to four. As the reference lattice, the Euclidean square lattice was taken. We
have used the RG-based algorithm, which approximates the ground state as the
product of identical tensors representing the congruent polygons; either squares
($p=4$) or pentagons ($p=5$) for the
Euclidean or hyperbolic lattices, respectively. We have analyzed the symmetries in
the tensors. The uniform TPS reduces the infinite number of the variational
parameters in the thermodynamic limit down to six (for $p=4$) or eight (for $p=5$) for the
Ising model in the presence of the transverse magnetic field. If no spontaneous
symmetry-breaking mechanism is present, the number of the free variational parameters
approximating the tensor product ground state further shrinks down being three only.
We have analyzed the phase transition of the Ising model by (1) the numerical calculation
of the optimized variational parameters, (2) the expectation value of the spin
polarization, (3) the phase transition magnetic exponent, (4) ground-state energy and
the magnetic susceptibility. The mean-field-like universality class was observed in the
vicinity of the phase transition magnetic fields $h_c^{(p)}$. The ground-state energy
per bond $E_0^{(p)}$ has been evaluated for the XY model and the Heisenberg model for
$p=4$ and $p=5$.

Our earlier studies of the classical spin lattice models on various types of the
hyperbolic surfaces~\cite{hctmrg-Ising-5-4,hctmrg-Ising-p-4,hctmrg-Ising-3-q} exhibited
a couple of interesting features, which are important to mention here. The classical
Ising model on the pentagonal lattice results in a higher phase transition temperature
if compared to the Euclidean square lattice~\cite{hctmrg-Ising-p-4}. Analogously, the
TFIM model in this work also exhibits the identical feature, i.e., $h_c^{(5)} >
h_c^{(4)}$. Classical spin systems on any hyperbolic lattice belong to the
mean-field-like universality class in accord with the current study. The
mean-field-like behavior of the exponents in the classical systems originates from the
hyperbolic lattice geometry, not from the numerical CTMRG method; the CTMRG accurately
reproduces the critical exponents on the 2D Euclidean lattices~~\cite{hctmrg-Ising-5-4,
hctmrg-Ising-p-4,hctmrg-Ising-3-q}. The low dimension of the tensors in the TPS
approximation of the {\it round-a-face} type suppresses the quantum long-range
correlations on the square lattice ($p=4$) near the criticality, where we obtained
$h_c^{(4)}$ deviating by $3.7\%$ if compared to the recent numerical analysis~\cite{HoSRG}.
The discrepancy is the consequence of the low dimension of the tensor $W_4$ appearing in
the ground-state approximation, which limits the numerical accuracy if approaching the
criticality, and it results in the mean-field exponent $\beta=\frac{1}{2}$. Therefore, it
is essential to distinguish between the mean-field approximation of the method used and
the mean-field-like behavior caused by the hyperbolic lattice geometry. That means that
we have dealt with both of the "mean-field" aspects in the current study.
The physical reasoning of the mean-field-like behavior caused by the hyperbolic lattice
comes from the exceedance of the critical dimensionality (being $d_c=3$ or $d_c=4$,
respectively, for the classical or quantum systems with the nearest-neighbor spin
couplings). Notice that the Hausdorff dimension of the pentagonal lattice is infinite.

Strictly speaking, classical spin systems exhibit the phase transitions exclusively
in the center of the infinite hyperbolic lattices because of strong boundary effects.
The ratio between the number of the spins on the boundary with respect to the number of
the spin on the remaining lattice area inside is $\approx2.73$ for the pentagonal lattice.
For this reason, the hyperbolic lattice becomes significantly sensitive to the boundary
effects~\cite{Baxter}. On the other hand, the boundary effects on the Euclidean lattices
play no role in the thermodynamic limit when the lattice size is expanded to infinity.
If calculating the correlation length $\xi$ on the pentagonal lattice (the distance
between two spins measured along a geodesics) for the classical Ising model, no divergence
at the phase transition was present, i.e., $0<\xi\lesssim1$~\cite{corrlen}. Moreover,
the correlation function was found to decay exponentially even at the phase transition,
which is the direct consequence of the finiteness of the correlation length at the phase
transition (and is related to an analogous exponential decay of the density matrix
spectra)~\cite{hctmrg-Ising-3-q}. In that sense, the phase transition on pentagonal
hyperbolic surface is non-critical, and we use the term {\it critical} for the spin
models on the square (Euclidean) lattice only since it is related to the divergence
of the correlation length by definition. The identical non-critical behavior for $p=5$
has also been confirmed in this work (for instance, by evaluating the non-divergent
magnetic susceptibility at $h_c^{(p)}$). It is highly non-trivial to evaluate the correlation
length unambiguously as both of the "mean-field" aspects are closely related and almost
impossible to be clearly separated or distinguished.

An exponentially fast decay of the reduced density matrix spectra has been observed for
the classical systems on hyperbolic surfaces~\cite{hctmrg-Ising-3-q,corrlen}. Similar
behavior for the quantum systems had to be taken into a possible scenario in the current
work. For all the above-mentioned reasons, we conjecture the TPFV analysis of the models
on the hyperbolic pentagonal lattice is more accurate than on the Euclidean ones. It is
caused by the mean-field-like behavior (the mean-field-like universality) as the consequence
of the pentagonal hyperbolic lattice with the infinite Hausdorff dimension, which has been
studied by the numerical improved mean-field approximation. Our preliminary results of the
identical quantum Hamiltonian on different hyperbolic lattices (to be published elsewhere)
support our claims and are in complete agreement with the conjectures we have made for the
classical spin systems.

\begin{acknowledgments}
We thank Tomotoshi Nishino and Frank Verstraete for valuable discussions. This work
was supported by the grants QIMABOS APVV-0808-12, VEGA-2/0130/15, and EU project SIQS No.
600645.
\end{acknowledgments}


\begin{thebibliography}{99}
\bibitem{rem} The Hamiltonian energy spectrum remains identical if $J_{xy}=\pm1$ provided
that $J_{z} = 0$ or $-1$ (for $h=0$) because the lattice is bipartite. Note, that
generalization to the antiferromagnetic ordering is trivial for the square geometry,
however, the pentagonal lattice geometry would result in strong frustration for all of
the models.
\bibitem{rem2} The non-physical (auxiliary) states in the TPS language are represented as
states with two degrees of freedom only. Such an approximation enhances the mean-field behavior
around the criticality and is a compromise to make the calculations feasible numerically
due to the exponentially increasing complexity of the hyperbolic lattice structure.
\end{thebibliography}
\end{document}